# Investigating the Impact of Arterial Irregularity On Clinical Parameters Using Reduced Order CFD Models In Stenosed Coronary Artery.

**Priyanshu Ghosh[1], Sayan Karmakar[2] , Disha Mondal[1],Oeshee Roy[1] and Supratim Saha [3],***

[1] Mechanical Engineering Department, Meghnad Saha Institute of Technology, Kolkata, India,
[2]Department of Civil Engineering, Jadavpur University, Kolkata, India.
[3] Department of Mechanical Engineering, Indian Institute of Technology Madras, Chennai, India.
*supratim@alumni.iitm.ac.in

**ABSTRACT**

Coronary heart disease (CHD) remains a leading cause of mortality worldwide. This study introduces a novel approach that integrates patient-specific Multi-slice CT scans into CAD models, using a one-dimensional numerical framework to assess varying degrees of coronary artery stenosis. The computational analysis encompasses the entire arterial tree, with a particular focus on stenosed coronary arteries modeled analytically. Key parameters, such as area and velocity, are derived from one-dimensional characteristic equations based on forward and backward characteristic variables. A resistance model with zero reflection coefficient and realistic pressure waveform inputs is applied at the outflow and inflow, respectively. The global characteristics captured by the 1D model serve as boundary conditions for a 2D axisymmetric model that focuses on local characteristics. The numerical solvers are validated against existing literature, ensuring grid independence. Fractional Flow Reserve (FFR) and Instantaneous wave-free Ratio (iFR) are calculated using various non-Newtonian models across different stenosis severities. The study also investigates the impact of lesion irregularity in stenosed coronary arteries, finding that irregular arteries exhibit lower FFR and iFR values and higher pressure drops, indicating increased blood flow resistance. This method provides a reliable, non-invasive diagnostic tool for evaluating the functional severity of irregular coronary artery stenosis in clinical settings, effectively capturing both global and local hemodynamic characteristics.

**Keywords**: Coronary artery, Stenosis, Irregularity, FFR , iFR,

## 1. INTRODUCTION

Coronary heart disease (CHD) is a significant global health concern, causing around 20 million deaths annually. In India, the CHD mortality rate rose from 17% (2001-2003) to 23% (2010-2013) [1]. Effective assessment methods for coronary artery disorders are essential for improving treatment strategies, particularly in severe cases requiring angioplasty or surgery. While mild CHD is generally treated with medication, many cases present a clinical challenge due to their intermediate severity, complicating treatment decisions. Traditional clinical assessments often rely on invasive techniques, which can be costly and unsuitable for managing medically treated stenosis [2]. Non-invasive alternatives, such as computed tomography angiography (CTA), provide 3D images of blood vessels and the heart, allowing for the estimation of clinical parameters like Fractional Flow Reserve (FFR). However, the computational demands of realistic stenosis modelling and blood vessel closure can be high. Previous studies have explored 3D numerical models for FFR estimation, but these are resource-intensive. In contrast, one-dimensional (1D) simulations present a more efficient approach while maintaining reasonable accuracy. Sherwin et al.[3] established key equations for blood flow in vessels, and Mynard and Nithiarasu [4] contributed relationships between systemic and coronary circulatory systems, which are integrated into a comprehensive 1D arterial network model in the present study. This study introduces a reduced-order computational fluid dynamics (CFD) model that utilizes patient-specific CT scan data to predict various clinical parameters, including FFR and instantaneous wave-free ratio (iFR). The innovation lies in modelling both global and local hemodynamic characteristics, employing a 1D model for boundary conditions alongside a 2D axisymmetric model for detailed local analysis. Comparative studies have shown the effectiveness of 1D versus 3D models in assessing hemodynamic parameters in coronary arteries [5,6]. Unlike previous work, which focused on 3D modelling, our method significantly reduces computational costs while maintaining accuracy. Additionally, various boundary conditions have been explored to compute hemodynamic parameters, as demonstrated in earlier studies [7,8,9]. This study aims to develop a reliable, non-invasive diagnostic tool for evaluating the functional severity of coronary artery stenosis by integrating patient-specific CT scan data into a 1D CFD model. This approach provides a quick, cost-effective alternative to traditional invasive techniques, enhancing clinical decision-making for CHD management. Addressing the critical parameter of lesion irregularity, it is seen that surface irregularity impact hemodynamic parameters, increasing the severity of blockages [10,11]. We have used non-Newtonian blood flow models, including the power-law, Carreau-Yasuda, and Casson models, which can better capture the complex rheological behaviour of blood [12]. The novelty of our approach lies in combining global and local hemodynamic characteristics through a 1D-2D coupled model, considering both non-Newtonian properties of blood and surface irregularity. This ensures a more accurate and detailed evaluation of stenosed coronary artery.

## 2. Methodology

### 2.1 Computational domain

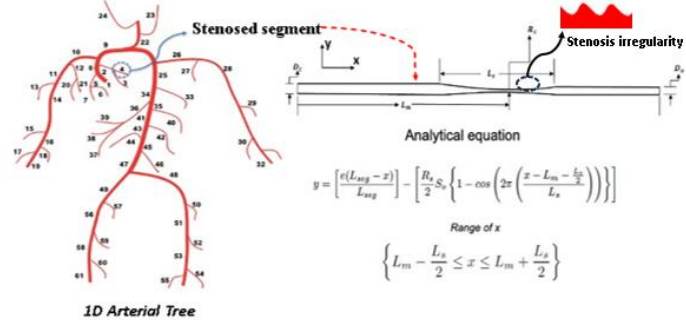

**Figure 1: Representative diagram of the computational domain.**

The computational domain for the 1D numerical simulation encompasses the full arterial tree having 61 segments, with the stenosed section (segment 4) specifically located in the left part of coronary tree, as illustrated in Figure 1. The stenosis is modelled using analytical equations that account for the tapering condition within the artery. Subsequently, the boundary conditions generated from the 1D arterial tree model are used in a higher-order 2D axisymmetric model. In this model, stenosis characterization relies on analytical formulations, incorporating periodic surface irregularities of the lesion. These irregularities are based on the height measurements obtained from a left circumflex coronary arterial cast, which exhibits mild, diffuse atherosclerotic disease [11]. For simplicity, the irregularity height is calibrated to this data, and the shape is assumed to follow a periodic pattern. The severity of stenosis of patient-specific cases are 40, 50 and 70 which are of intermediate grade stenosis. This hierarchical modelling approach allows for a comprehensive analysis, effectively capturing both the global characteristics of the entire arterial network and the local hemodynamic details within the stenosed region.

#### 2.1.1 1D numerical details.

The vessel is depicted as a shape approximating a cylindrical vessel with a compliant surface on the wall where blood will be flowing. In the study by Sherwin et al. [3], the one-dimensional equations for mass and momentum conservation are presented.

$$\frac{\partial A}{\partial t} + \frac{\partial (Au)}{\partial x} = 0 \qquad (1)$$

$$\frac{\partial u}{\partial t} + u\frac{\partial u}{\partial x} + \frac{1}{\rho}\frac{\partial p}{\partial x} - \frac{f}{\rho A} = 0 \qquad (2)$$

In this model, the area of cross-section is denoted as $A$, the velocity taken after averaging across the flow area is $u$, and $p$ represents the pressure within the vessel where blood is flowing. The density ($\rho$) of the blood is assumed to be 1060 kg/m³, and ($f$) denotes the frictional force. The assumptions of Steady and laminar flow conditions, indicative of Poiseuille flow, are assumed for modeling the friction term [4]. The system of equations is closed by introducing supplementary constraints that relate the pressure to the vessel's cross-sectional area. The relationship between area and pressure depends on factors such as elasticity (wall behaviour), thickness of wall, and Poisson's ratio, as discussed in the literature [4,5,13].

$$p = p_{ext} + \beta\,(\sqrt{A} - \sqrt{A_0}) \qquad (3)$$

Where $p_{ext}$ represents the pressure which is acting trasmurally, $A_0$ is the area of cross-section when the transmural pressure is zero ( i.e. $p = p_{ext}$ ), and ( $\beta$ ) denotes the properties of the blood vessel material.

$$A = \frac{(W_1 - W_2)^2}{1024}\left(\frac{\rho}{\beta}\right)^2 \qquad (4)$$

$$u = \frac{1}{2}(W_1 + W_2) \qquad (5)$$

The unknown parameters, such as velocity ($u$) and area ($A$), are determined using traveling characteristics (forward and backward), as shown in Equations 4 and 5. In these equations, $W_1$ and $W_2$ denote the properties of waves that are propagating in the forward and backward directions, respectively. The numerical simulation is performed by means of the Locally Conservative Galerkin (LCG) method [4, 5,13].

### 2.1.3 Boundary conditions.

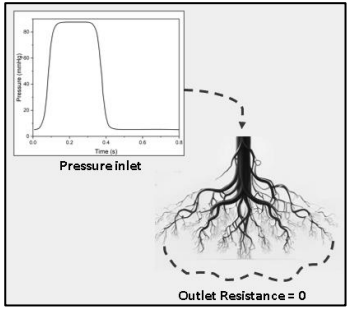

**Figure 2: 1D boundary condition implemented in arterial tree**

As illustrated in Figure 2, the representative diagram displays the conditions which are imposed on boundary in the 1D numerical model. A boundary condition related to pressure, generated using a sigmoid function, is imposed at the beginning of the arterial tree. At the outlet, a zero reflection coefficient is assigned. Detailed information about the conditions which is prescribed on boundary using characteristic variables can be found in the literature [4].

### 2.1.4 Grid independence study

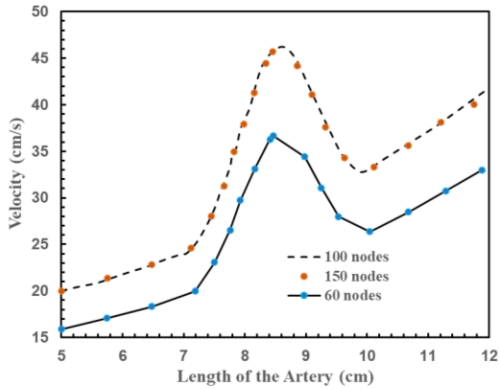

**Figure 3: 1D grid independent study of axial velocity**

For a specific severity scenario, numerical analysis is performed using three grid sizes: 50, 100, and 150 grid points. The magnitude of velocity in the artery's flow direction is determined and compared across these mesh sizes, as shown in Figure 3. The mesh configuration having 100 and 150 grid points exhibit similar trends. Therefore, in the numerical simulation using the configuration of 100 grid points per segment of the arterial tree is used.

### 2.1.4 Validation study

The present results by current solver, shown in Figure 4, are compared with the numerical results reported by Low et al. [15]. The right carotid artery was selected for the validation study. The observed trends in both pressure and flow align well under normal conditions and during cardiac function.

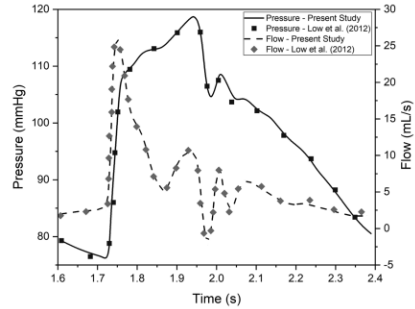

**Figure 4: Pressure and flow waveform comparison with literature result for Right Carotid artery [15]**

### 2 2D Numerical details

Given that blood flow through the arterial vessel is considered to be laminar and incompressible, the mass and momentum conservation equations are as follows:

$$\frac{\partial \rho}{\partial t} + \nabla \cdot (\rho V) = 0$$

$$\rho \frac{\partial V}{\partial t} + \rho (V.\nabla)V = \nabla.[-pI + \mathbf{T}] + F$$

$$Where, \; \mathbf{T} = \mu_{app}(\nabla.V + (\nabla.V)^T)$$

The Blood is modelled as Newtonian and non-Newtonian model like Carreau Yesuda, Casson model and power law model. The apparent viscosity ($\mu_{app}$) is different according to the model considered. The coefficient of the models are taken from literature [12].

### 2.3 Boundary Conditions.

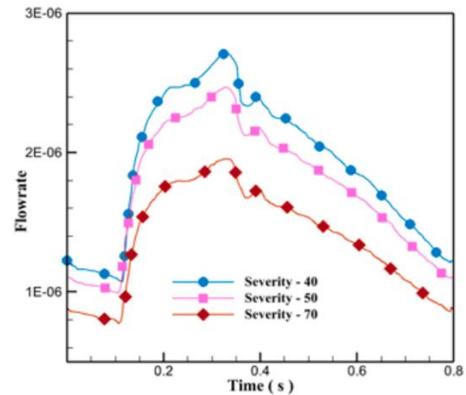

**Figure 5: 1D boundary condition generated for different severity which is given as inlet to higher order model**

The boundary conditions for 2D axisymmetric model are generated using 1D model for flowrate and pressure conditions. At inlet, flowrate is prescribed and at the outlet, pressure is imposed. The flowrate for different severity obtained from 1D model are shown in Figure 5. Table 1 shows the pressure conditons which are imposed at the outlet.

**Table 1: Pressure boundary condition generated from 1D model**

| Sr no | Severity ( % ) | Outlet pressure (mmHg) |
|---|---|---|
| 1 | 40 | 106.154 |
| 2 | 50 | 104.296 |
| 3 | 70 | 93.303 |

### 2.2.4 Grid independence study

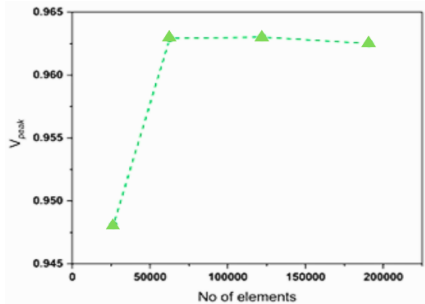

**Figure 6: Peak velocity variation with no of elements**

A grid independence analysis is conducted for all cases involving stenosis but only one case for peak velocity ($V_{peak}$) is shown in Figure 6. The error is assessed in relation to the highest resolution mesh used in the simulation. The numerical simulation is performed in the computational domain using around 200000 elements to ensure grid independent solution.

### 2.2.5 Validation study

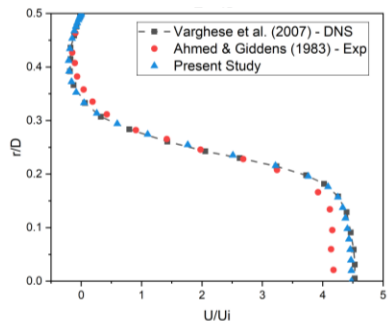

**Figure 7: Velocity comparison with literature result [15, 16]**

The blood flow through an occluded artery using 2D axisymmetric numerical model is validated against the literature results [15,16]. The Reynolds number (Re) for the flow is based on the characteristic length (D) and reference velocity ($U_{avg}$) at the inlet, is defined as (Re = $\rho U_{avg} D/\mu$ ). The system level parameters used in the simulation are diameter (D = 1 m), viscosity ($\mu$ = 1 Pa·s), velocity ($U_{avg}$ = 0.5 m/s), and density ($\rho$ = 1000 kg/m³). Diameter and velocity are non-dimensionalized. At the inlet, velocity profile having parabolic variation is imposed, with a no relative motion in the tangential direction on the wall and zero pressure (p = 0) at the outlet [18,19]. The velocity is compared at a locations (1D) downstream of stenosed section as shown in Figure 8, matches well with the literature.

## 3. Result and discussion

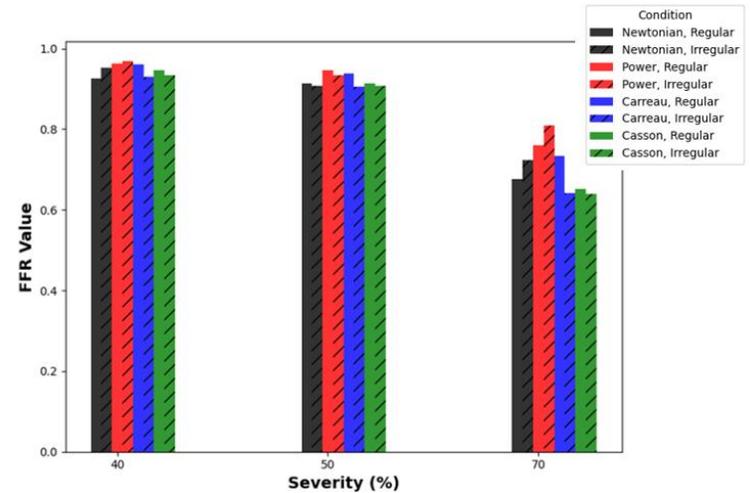

**Figure 8: Comparison of FFR values for different smooth and irregular stenosed coronary arteries**

Figure 8 presents a comparison of Fractional Flow Reserve (FFR) values across varying levels of arterial blockage severity, using different fluid models, and contrasting smooth versus irregular artery conditions. The irregular arteries exhibit lower FFR values compared to smooth arteries, suggesting that irregularities in the arterial wall contribute to greater resistance to blood flow, leading to a more significant drop in FFR. This trend is consistent across different severity levels, where increased severity typically correlates with lower FFR values, indicating a higher degree of blood flow restriction. The literature [10] shows that irregularities in the artery affect hemodynamic parameters, such as Wall Shear Stress (WSS). While our study is focused on Fractional Flow Reserve (FFR), we find that the impact of arterial irregularities on FFR aligns with these observations, indicating a consistent relationship between arterial irregularities and hemodynamic behavior. Additionally, it also reveals the differences in FFR predictions across the fluid models, emphasizing the impact of selecting an appropriate model to accurately assess coronary artery disease. The Carreau-Yasuda model, which includes both shear-thinning and viscoelastic properties, reveals slight differences in FFR values, indicating its sensitivity to subtle changes in blood flow dynamics and its capability to accurately simulate blood flow under a wide range of shear rates. Similarly, the Casson model effectively captures the yield stress behavior of blood, showing slight differences in FFR values and highlighting its importance for understanding flow in microcirculatory systems and low shear conditions.

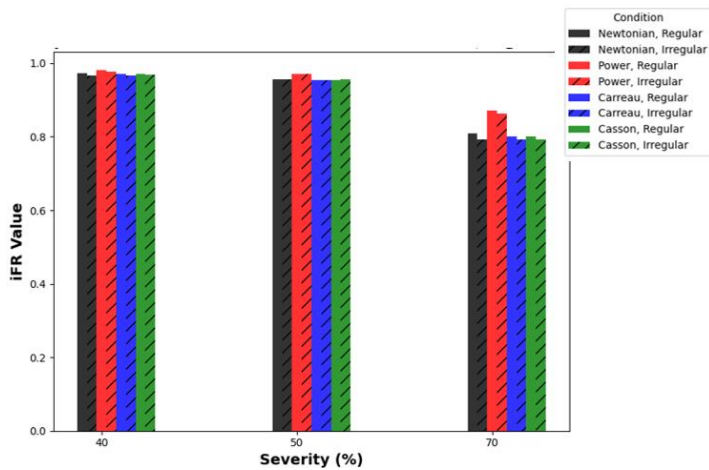

**Figure 9: Comparison of iFR values for different smooth and irregular stenosed coronary arteries.**

Figure 9 compares the Instantaneous wave-Free Ratio (iFR) values across varying severity levels of arterial blockage, different fluid models, and between smooth and irregular artery without the need for the hyperemic agents. It demonstrates that, generally, iFR values decrease as the severity of stenosis increases, reflecting greater blood flow restriction with more significant arterial blockage. Moreover, irregular arteries tend to have slightly lower iFR values compared to smooth arteries for the same model and severity, indicating that irregularities in the arterial wall contribute to increased resistance to blood flow. The Newtonian model shows a slight difference in iFR values between smooth and irregular stenosed arteries, indicating some sensitivity to vessel geometry changes. The Power Law model, again, shows minimal differences in iFR values, which may not able to capture blood flow dynamics during the wave-free period due to irregularity. The Carreau-Yasuda model shows a very slight difference in iFR values. The Casson model captures the initial resistance to flow due to yield stress, showing slight differences in iFR values, although it may be less applicable for larger vessels under high shear rates. Figure 10 illustrates the comparison of pressure drop values across different severity levels of arterial blockage, various fluid models, and between smooth and irregular artery conditions. It shows that, for each severity level, irregular arteries generally experience a higher pressure drop compared to smooth arteries under the same conditions. This indicates that the irregularities in the arterial wall contribute to increased resistance, leading to a more significant pressure drop.

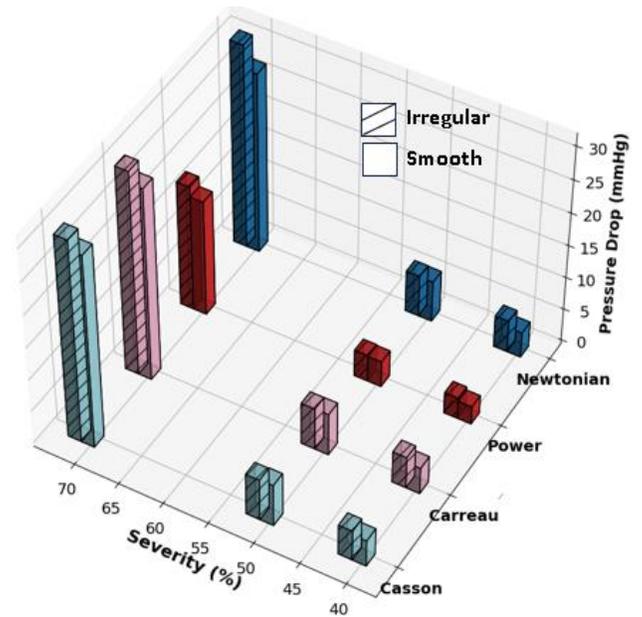

**Figure 10: Comparison of pressure drop values different smooth and irregular stenosed coronary arteries.**

Additionally, as the severity of stenosis surges, the pressure drop also increases, particularly in the more complex non-Newtonian models like Carreau and Casson. This information is crucial for understanding the hemodynamic effects of coronary artery disease and can inform clinical decisions regarding the assessment and management of arterial blockages. Smooth stenosed arteries tend to exhibit slightly higher FFR and iFR values, indicating lower resistance to flow, while irregular stenosed arteries generally show higher pressure drops across most models, indicating greater resistance to flow. Each rheological model provides unique insights, the Newtonian model highlights geometric differences but lacks complex flow condition accuracy. the Power Law model offers robust predictions by accounting for shear-thinning properties. The Carreau-Yasuda model provides a comprehensive view of blood flow dynamics with increased computational complexity and the Casson model is ideal for modeling yield stress effects, particularly in microcirculation. The choice of rheological model should be guided by specific clinical or research requirements, balancing accuracy, computational efficiency, and the complexity of blood flow dynamics being studied. These results underscore the importance of incorporating non-Newtonian models for more precise predictions of blood flow in stenosed coronary arteries, which is essential for informed clinical decision-making and intervention planning. This approach can help prevent both overtreatment and undertreatment. However, it is crucial to apply these models with caution to ensure balanced and effective outcomes.

## CONCLUSION

This study demonstrates that arterial wall irregularities and stenosis severity significantly impact hemodynamic parameters, including Fractional Flow Reserve (FFR), Instantaneous wave-Free Ratio (iFR), and pressure drop values. Irregular arteries exhibit lower FFR and iFR values and higher pressure drops, indicating increased blood flow resistance. Various rheological models are employed in the simulation, with each model emphasizing different aspects of the flow. A comparison of clinical parameters is conducted using these models. These findings highlight the importance of selecting appropriate fluid models to accurately assess coronary artery stenosis. Incorporating non-Newtonian properties into computational models enhances the precision of diagnostic tools, leading to better clinical decision-making and improved patient outcomes.

## ACKNOWLEDGEMENTS

The corresponding author expresses gratitude for the PMRF scheme and the associated funding (SB22230230MEPMRF008509), which have significantly contributed to advancing knowledge sharing among undergraduate students in private engineering colleges across India. This support is pivotal in closing research gaps and cultivating a research-driven mindset among the next generation of scholars.

## REFERENCE

1. Gupta R, Mohan I, Narula J. Trends in coronary heart disease epidemiology in India. Annals of global health.82(2),307-15 (2016).
2. Koo, B. K., Erglis, A., Doh, J. H., et al. (2011). Diagnosis of ischemia-causing coronary stenoses by noninvasive fractional flow reserve computed from coronary computed tomographic angiograms. Journal of the American College of Cardiology, 58(19), 1989-1997.
3. Sherwin, S. J., Formaggia, L., Peiro, J., & Franke, V. (2003). Computational modelling of 1D blood flow with variable mechanical properties and its application to the simulation of wave propagation in the human arterial system. International Journal for Numerical Methods in Fluids, 43(67), 673-700.
4. Mynard, J. P., & Nithiarasu, P. (2008). A 1D arterial blood flow model incorporating ventricular pressure, aortic valve, and regional coronary flow using the locally conservative Galerkin (LCG) method. Communications in Numerical Methods in Engineering, 24(5), 367-417.
5. Saha, S., Purushotham, T., & Prakash, K. A. (2020). Comparison of fractional flow reserve value of patient-specific left anterior descending artery using 1D and 3D CFD analysis. International Journal of Advances in Engineering Sciences and Applied Mathematics.
6. Boileau, E., Nithiarasu, P., Blanco, P. J., Müller, L. O., Fossan, F. E., Hellevik, L. R., Donders, W. P., Huberts, W., Willemet, M., & Alastruey, J. (2015). A benchmark study of numerical schemes for one-dimensional arterial blood flow modelling. International Journal for Numerical Methods in Biomedical Engineering, 31(10), e02732.
7. Bit, A., & Chattopadhyay, H. (2014). Numerical investigations of pulsatile flow in stenosed artery. Acta of Bioengineering and Biomechanics, 16(4).
8. Coccarelli, A., Saha, S., Purushotham, T., Prakash, K. A., & Nithiarasu, P. (2021). On the poro-elastic models for microvascular blood flow resistance: An in vitro validation. Journal of Biomechanics, 117, 110241.
9. Saha, S., Purushotham, T., & Prakash, K. A. (2019). Numerical and experimental investigations of Fractional Flow Reserve (FFR) in a stenosed coronary artery. E3S Web of Conferences, Vol. 128. EDP Sciences.
10. Ellis, Stephen, et al. "Morphology of left anterior descending coronary territory lesions as a predictor of anterior myocardial infarction: a CASS Registry Study." Journal of the American College of Cardiology 13.7 (1989): 1481-1491.
11. Chakravarty, Santabrata, and Prashanta Kumar Mandal. "Effect of surface irregularities on unsteady pulsatile flow in a compliant artery." *International Journal of Non-Linear Mechanics* 40.10 (2005): 1268-1281.
12. Shibeshi, Shewaferaw S., and William E. Collins. "The rheology of blood flow in a branched arterial system." Applied rheology 15.6 (2005): 398-405.
13. Saha, S., Purushotham, T., & Prakash, K. A. (2021). Effect of Occlusion Percentage and Lesion Length on Stenosed Coronary Artery: A Numerical Study. In Recent Advances in Computational Mechanics and Simulations (pp. 87-97). Springer, Singapore.
14. Saha, Supratim. "Numerical investigation of blood flow through stenosed coronary artery using reduced order model." Conference on Fluid Mechanics and Fluid Power. Singapore: Springer Nature Singapore, 2021.
15. Low K, van Loon R, Sazonov I, Bevan RL, Nithiarasu P. An improved baseline model for a human arterial network to study the impact of aneurysms on pressure- flow waveforms. International journal for numerical methods in biomedical engineering.28(12),1224-46(2012).
16. Varghese, Sonu S., Steven H. Frankel, and Paul F. Fischer. "Direct numerical simulation of stenotic flows. Part 1. Steady flow." Journal of Fluid Mechanics 582 (2007): 253-280.
17. Ahmed, Saad A., and Don P. Giddens. "Velocity measurements in steady flow through axisymmetric stenoses at moderate Reynolds numbers." *Journal of biomechanics* 16.7 (1983): 505-516.
18. Karmakar, S., Pal, A., Sarkar, S., & Mukhopadhyay, A. (2025). Optimized designs for high-efficiency particle sorting in serpentine microfluidic channels. *Physics of Fluids, 37*(4), 042006.
19. S. Karmakar, M. S. Mondal, A. Pal, and S. Sarkar, "Optimizing sorting of micro-sized bio-cells in symmetric serpentine microchannel using machine learning," *arXiv preprint arXiv:2308.01701*, 2023. [Online].